\documentclass[twocolumn, amsmath,superscriptaddress, nohypertex, amsfonts,prb]{revtex4}
\usepackage{graphicx}
\usepackage{url}
\pdfoutput=1
\begin{document}

\title{Percolation with plasticity for neuromorphic computing}

\author{V. G. Karpov}
\email{victor.karpov@utoledo.edu}
\affiliation{Department of Physics and Astronomy, University of Toledo, Toledo, OH 43606, USA}
\author{Maria Patmiou}
\email{maria.patmiou@rockets.utoledo.edu}
\affiliation{Department of Physics and Astronomy, University of Toledo, Toledo, OH 43606, USA}

\date{\today}

\begin{abstract}

We introduce the percolation with plasticity (PWP) systems that exhibit neuromorphic functionalities including multi-valued memory, random number generation, matrix-vector multiplication, and associative learning. PWP systems have multiple ($N\gg 1$) interfaces with external circuitry (electrodes) allowing $N!\gg 1$ measurable interelectrode resistances. Due to the underlying material properties, they undergo successive nonvolatile modifications in response to electric pulses. PWP networks offer some advantages over the existing neural network architectures. Overall, random self-tuning PWP systems with high degree of parallelism, multiple inputs and outputs present close similarities to the cortex of mammalian brain.  Understanding their topology, electrodynamics, and statistics opens a field of its own calling upon new theoretical and experimental insights.

\end{abstract}
\maketitle
\section{Introduction}\label{sec:intro}

Devices for neuromorphic computing remain among the most active ares of research in computer science and engineering, and artificial intelligence with a variety of models for neurons, synapses and their networks. \cite{schuman2017,zidan2018,bohnstingal2019,sebastian2019,wan2019,zhang2019,upadhhyay2019} Such devices are typically built of nonvolatile memory cells and interconnects  wired in a certain architecture.

Here we introduce a concept of neuromorphic devices where no artificial memory cells and interconnects are required. They are based on disordered  materials with percolation conduction  \cite{efros,shik,snarskii} including amorphous, polycrystalline and doped semiconductors, and granular compounds. We recall that the percolation conduction takes place in systems of microscopic exponentially random resistors and is dominated by the infinite cluster built of multiple smallest resistors allowing electric connectivity.

Of all possible percolation materials, here we consider those exhibiting plasticity, i. e capable of changing their resistances in response to electric bias. They include metal oxides and chalcogenide compounds used respectively with resistive random access memory (RRAM) \cite{lanza2014} and phase change memory (PCM), \cite{sebastian2019} granular metals, \cite{gladskikh2014} and nano-composites.\cite{song2016}

\begin{figure}[t]
\includegraphics[width=0.32\textwidth]{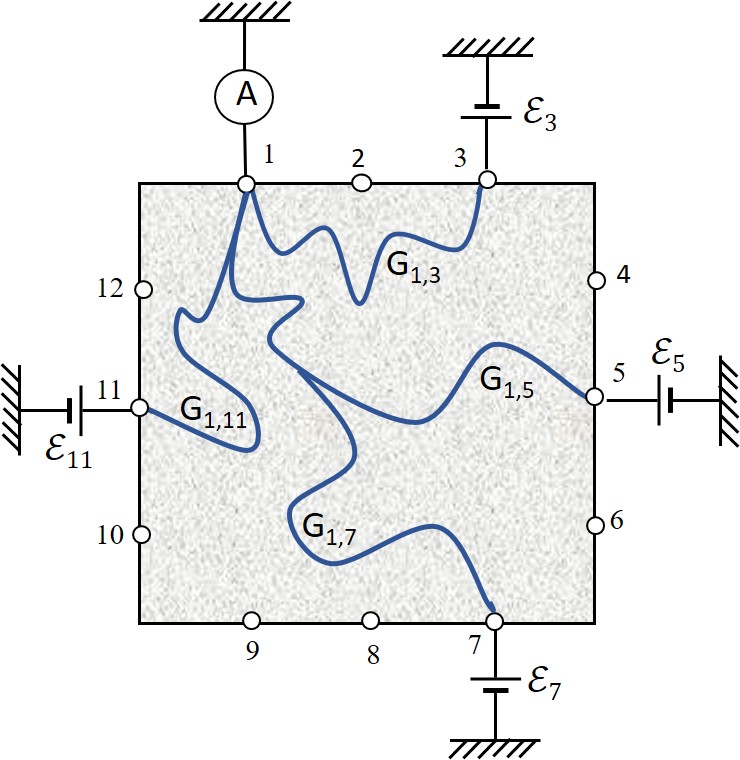}
\caption{Schematic 2D illustration of percolation systems with multiple local interfaces (electrodes). $G_{ij}$ stand for pathway conductances. Only a small number of $12!\approx 10^{13}$ pathways possible in the sketch is shown.\label{Fig:PWPconcept}}
\end{figure}

Such percolation with plasticity (PWP) systems are illustrated in Fig. \ref{Fig:PWPconcept} showing some of the interelectrode pathways and assuming a relatively small number of electrodes. Anticipating a particular application below, Fig. \ref{Fig:PWPconcept} assumes certain voltages ${\cal E}_i$ applied to all electrodes but one used to measure the electric current $I$. Other implementations would allow different circuitry.

We note the following features of PWP systems.\\ (i) The exponentially large number $M=N!\approx \exp(N\ln N)\gg 1$ of interelectrode resistances $R_{ij}$ that scales exponentially with the number $N\gg 1$ of electrodes. For example, $M\sim 10^{13}$ in a design of Fig. \ref{Fig:PWPconcept}. \\
(ii) Multivalued memory in conductive pathways operated by electric pulses that modify $R_{ij}$ due to plasticity; \cite{patmiou2019} they play the role of microscopic memory cells. \\
(iii) Direct connectivity between the bond-forming microscopic resistors eliminates the need for artificial interconnects.\\
(iv) The multivalued memory in combination with multiplicity $M\gg 1$ offers a platform for in-memory computing. \cite{verma2019} Furthermore, mathematically, series of cells in PWP present multidimensional random vectors forming a base for hyperdimensional computing. \cite{kanerva2009,mitrokhin2019,karunaratne2019}\\
(v) The randomness of PWP topology offers a natural implementation to the randomly wired neural networks outperforming their regularly wired counterparts. \cite{hie2019,zopf2018}

In what follows we consider the statistics of resistances and switching characteristics of PWP pathways and present examples of the neuromorphic functionality.

\section{PWP vs. standard percolation}\label{sec:PWPST}
(1) {\it Standard percolation}\cite{efros,shik,snarskii} is dominated by the sparse infinite cluster between infinitely large area electrodes on the opposite faces of a sample. That cluster's bonds consist of the minimally strong resistors with total concentration sufficient to form a connected structure. It is effectively uniform over large distances $L\gg L_c$ where $L_c$ is the correlation radius determining the characteristic mesh size of the cluster.

Each bond of the cluster consists of a large number ($i=1,2,..$) of exponentially different random resistors, $R_i=R_0\exp(\xi _i)$ where random quantities $\xi _i$ are uniformly distributed in the interval $(0,\xi _m )$. $\xi _m$ can be determined by the requirement that resistors with $\xi <\xi _m$ form the infinite cluster.

The physical meaning of $\xi$ depends on the type of system. For definiteness, we assume here $\xi _i=V_i/kT$ corresponding  to random barriers $V_i$ in noncrystalline materials where $k$ is the Boltzmann's constant and $T$ is the temperature. In reality, the nature of percolation conduction can be more complex including e. g. finite size effects and thermally assisted tunneling between the microscopic resistors in nanocomposites. \cite{lin2013,eletskii2015} These complications will not qualitatively change our consideration here.

The microscopic resistors of a percolation cluster exhibit non-ohmicity due to the field induced suppression of their barriers $V_i$,
\begin{equation}\label{eq:nonohm}J_i=J_0\exp(-V_i/kT)\sinh(qU_i/2kT)\end{equation}
where $J$ is the resistor current, $U_i=E_ia$ is the voltage applied to the barrier, $a$ and $E_i$ are respectively the barrier width and local electric field, and $q$ is the electron charge.
\begin{figure}[b!]
\includegraphics[width=0.32\textwidth]{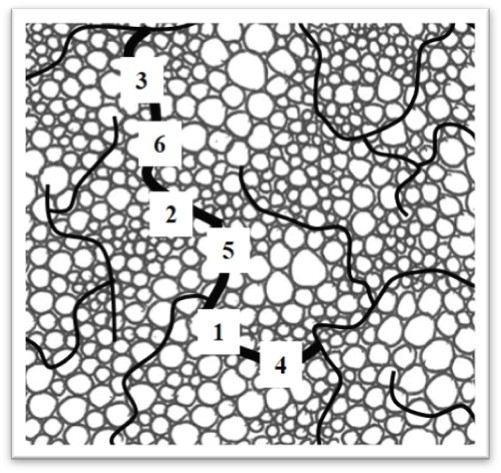}
\caption{A fragment of conductive pathways in the infinite percolation cluster representative of polycrystalline or granular materials. Numbers 1-6 represent random resistors in descending order. \label{Fig:PWP2}}
\end{figure}

An important conceptual point by Shklovskii \cite{shklovskii1979,aladashvili1989,patmiou2019} is that the applied voltage concentrates on the strongest resistor of a percolation cluster bond (resistor 1 in Fig. \ref{Fig:PWP2}) suppressing it to the level of the next strongest (resistor 2 in Fig. \ref{Fig:PWP2}), so the two equally dominate the entire bond voltage drop. Along the same lines, it then suppresses the next-next strongest resistors (3,4,5,.. in Fig. \ref{Fig:PWP2}), etc. As a result, the percolation cluster changes its structure,
\begin{equation}\label{eq:corrad}L_c=a\sqrt{V_0/qEa},\quad  \Delta V=\sqrt{V_mqaE}\end{equation}
resulting in the macroscopic non-ohmic conductivity, $\sigma =\sigma _0\exp\left(\sqrt{V_mqaE}/kT\right)$
where $L_c$ and $\Delta V$ are respectively the field dependent correlation radius and maximum barrier decrease in the percolation cluster, and $V_0$ is the amplitude of barrier variations.

Two assumptions underly the latter theory: (a) The volatility of bias induced changes: each microscopic resistor adiabatically following the bias. (b) The quasistatic nature of biasing with times $t\gg \tau _m=\tau _0\exp(\xi _m)$ where $\tau _0\sim 1$ ps is on the order of the characteristic atomic vibration times, limiting applications to pulse biasing and invoking various transients.

(2) {\it PWP} differs from the standard percolation in both topology and non-ohmicity. The former is such that the hereby proposed PWP has multiple ($N\gg 1$) electrodes; the latter is due to the nonvolatile nature of bias induced changes.

The concept of infinite cluster survives if the electrode sizes $l$ and interelectrode distances $L_{ij}$ are much larger than $L_c$, in which case (we call it `large PWP')  $R_{ij}$ are determined by geometry, $R_{ij}=(\sigma l)^{-1}f_{ij}(l/L_{ij})$. Here $f_{ij}$ is a dimensionless function whose shape depends on the electrode locations through the confinement of electric currents by the sample boundaries. While $f_{ij}$ can be numerically modeled with for any particular electrode configurations, some general statements can be made based on the available examples. \cite{hong2019} Using a rough model of the electrode as an immersed metal hemisphere, a rough estimate $f_{ij}\approx 1+O(l/L_{ij})+O(l/L)$ emerges, where $O(x)$ means `of the order of $x$'. Because the number $M$ of different $L_{ij}$s is exponentially large, the spectrum of resistances is narrow and dense, with the characteristic variations of the order of
\begin{equation}\label{eq:deltaR}\delta R\sim \langle{R_{ij}\rangle}/M\ll \langle{R_{ij}\rangle} \end{equation} where angular brackets represent averaging. The macroscopic conductivity $\sigma$ in the equation for $R_{ij}$ is taken in the limit of infinitely large percolation systems where it is uniquely defined. The system finiteness will slightly smear the corresponding predictions introducing a relative dispersion \cite{efros} $\sim L_c/L\ll 1$.

The concept of infinite cluster fails when $l$ and/or $L_{ij}$ are comparable with or smaller than $L_c$. In the latter (`small PWP') case, one has to consider multiple conductive paths unrelated to the infinite cluster. Given a broad variety of different percolation systems both the cases of large and small PWP are possible. For example, the modern $\sim 10$ nm- node technology corresponds to the case of small PWP given \cite{patmiou2019} $a\gtrsim 0.3$ nm and $V_0/T\lesssim 100$ i. e. $L_c\sim 30$ nm. Increasing $l$ towards the micron node systems will result in large PWP networks.

The important non-ohmicity feature of PWP is its nonvolatile nature rendered by the underlying material (say, of PCM or RRAM type). We assume that each microscopic element of a conductive path can exist in either  high or low-resistive state whose respective resistances, $R_{>}$ and $R_{<}$ are markedly different. The applied bias concentrated on the strongest resistor (in the manner of Fig. \ref{Fig:PWP2}) will change it from $R_{>}$ to $R_{<}$ {\it by switching}, i. e. by long lived structural transformation not adaptable to subsequent voltage variations. Hence, the next strongest resistor will be stressed with practically {\it the same voltage} as opposed to the above case of volatile non-ohmicity. In the first approximation, an originally resistive  percolation bond will transform into its conductive state by $n$ discrete steps where $n$ is the number of its microscopic resistances, similar to a falling row of dominoes arranged in the order of descending $\xi$'s.

Two comments are in order with regards to PWP switching. First, it requires strong enough electric fields $E>E_c$ where the critical field $E_c$ decreases logarithmically with the electric pulse (spike) length. \cite{karpov2008,krebs2009,bernard2010,sharma2015,you2017} That temporal dependence (along with the above mentioned transients) opens a venue to the spike timing dependent plasticity (STDP), \cite{markram2011} which is another important property of neural networks. Note that because almost the entire voltage drops across a microscopic resistance of small linear size $a$, the microscopic field $E$ is stronger than the apparent macroscopic field by the factor of $L_c/a\gg 1$ or $L_{ij}/a\gg 1$ for the cases of respectively large and small PWP structures. The ratio $L_{ij}/a\gg 1$ will vary significantly between the pathways of different lengths (such as shown in Fig. \ref{Fig:PWPconcept}); hence the possibility of targeted memory records purposely using designated pairs of electrodes with small PWP devices.

More accurately, for a chain of $n$ resistors $R_{>,i}$ arranged in descending order, the field on the highest resistor left after $m$ $(<n)$ switchings is given by
\begin{equation}\label{eq:field}E_m=E_{\rm max}R_{>,m}[R_{>,m}+\sum _{m+1}^nR_{>,i}+(m-1)R_{<})]^{-1}\end{equation}
where $E_{\rm max}$ is the maximum field on the maximum resistor before switching started. Starting with $E_{\rm max}$ only slightly above $E_c$, the sequence of switching will cease (or significantly slow down) after a number of events when $E_m<E_c$. Further increasing the field will resume that process creating multiple records per chain.

Secondly, the switching process can be terminated in case of close enough microscopic resistances sharing almost equally the applied voltage and thus insufficiently strong field $E<E_c$ each. While such coincidences seem unlikely requiring two or more random quantities $\xi$ on a bond falling in the same interval $\delta\xi\lesssim 1\ll \xi _m$, they can lead to switching failures calling upon future work.

Finally, we give a rough estimate of the statistics of resistances for small PWP systems. Based on Eq. (\ref{eq:nonohm}) we start with an approximation for a resistance of a chain, $R=R_{\rm max}\exp(-\delta V/kT)$ where $\delta V=V_{\rm max}-V$ and $V$ is the maximum barrier in that chain, $R_{\rm max}=R_0\exp(V_{\rm max}/kT)$, and $V_{\rm max}$ is the maximum barrier in the entire system. In this approximation, the effect of a number of microscopic resistors in a chain is relatively unimportant compared to the exponential effect of random barriers. Assuming uniformly distributed barriers, for a $n$-resistor chain, the average number of resistors with the barriers above a given $V$ is $n_V=n(V_{\rm max}-V)/(V_{\rm max}-V_{\rm min})$ where $V_{\rm min}$ is the minimum barrier. The probability of having no barriers greater than $V$ is $P_V(n)=\exp(-n_V)$. To further simplify the problem we assume $n\gg 1$ so that $nkT/(V_{\rm max}-V_{\rm min})>1$, i. e. $P_V(n)$ strongly decays beyond the significant interval of $kT$ near $V$; hence, $P_V(n)kT$ gives the probability of finding the chain barrier within $kT$ around $V$.

Multiplying $P_V(n)$ by the probability $P_n(L)\sim (L/an)\exp(-L^2/na^2)$ of finding a n-resistor chain connecting points distance $L$ from each other, we obtain the probability density of n-chain with a given barrier $V$ (to the accuracy of a numerical multiplier in the exponent). Because in the approximation employed the resistance $R$ does not depend of $n$, we integrate the latter product over $n$ and express $V$ through $R$, which yields the probabilistic distribution density,
\begin{equation}P(R)\propto \frac{1}{R}\exp\left(-\frac{2L}{a}\sqrt{\frac{kT}{V_{\rm max}-V_{\rm min}}\ln\frac{R_{\rm max}}{R}}\right).\end{equation}
It follows that resistance spectrum is a gradual function with a minimum at $\ln (R_{\rm max}/R)=(L/a)^2kT/V$ and the characteristic interval between two subsequent values obeying the inequality of Eq. (\ref{eq:deltaR}).

Similar to the standard percolation theory, the above analysis ignores possible transients that may be important for the case of electric pulses. \cite{callaghan2018}  Neither it addressed the important question of reversibility of switched structures presenting another challenge. Two conceivable scenarios include thermally induced transitions from low- to high-resistive states, as well as implementation of materials with a degree of ferroelectricity allowing reversibility in response to electric polarity changes. \cite{karpov2017}

\section{Examples of functionality}
{\it Multivalued memory.} When a certain bias is applied between any two electrodes, their connecting path will change its resistance to a random, but unique value. Given for example $N=10$ electrodes on the faces of 1x1x1 cm$^3$ cube, the number of perceptive pathways $M\sim 10^{10}$  cm$^{-3}$ is higher than that of human cortex. The same 3D 1x1x1 cm$^3$ structures with multiple electrodes at each side will result in a higher memory capacity than the current 3D crossbar architecture. It will be further enhanced with the functionality of multiple records per one electrode pair.

{\it Generation of random numbers.} Lets show that the measurable variations $\Delta R_{ij}$ in the interelectrode resistances $R_{ij}$ are random quantities mutually uncorrelated with any desired accuracy. Consider  $\Delta R_{ij}=R_{ij}-\langle R_{ij}\rangle$ with $R_{ij}= R_0\sum _1^{N_{ij}}\exp(\xi _i)$ for a bond of $N_{ij}$ resistors and $\xi _i$ uniformly distributed in the interval $(\xi _{\rm min},\xi _{\rm max})$. It is then straightforward to obtain the correlation coefficient between the resistances of $(i,j)$ and $(k,l)$ bonds,
\begin{equation}\label{eq:corco}
C\equiv \frac{\langle\Delta R_{ij}\Delta R_{kl}\rangle}{\sqrt{\langle(\Delta R_{ij})^2\rangle\langle(\Delta R_{kl})^2\rangle}}=\frac{N_s}{\sqrt{N_{ij}N_{kl}}}
\end{equation}
where $N_{ij}$ and $N_{kl}$ represent the numbers of microscopic resistances in those bonds, and $N_s$ is the number of resistances shared between them. We describe each bond as a random walk. Then, if the two bonds are not close geometrically, separated by distances $L_{ijkl}$ exceeding $a\sqrt{N_{ij}+N_{kl} }$, then their overlap is exponentially small, and $C\sim \exp\{-L_{ijkl}^2/[a^2(N_{ij}+N_{kl})]\}\ll 1$. The averages implied by the definition for $\Delta R_{ij}$ can be readily measured for an ensemble of geometrically similar pairs, such as (1,9), (2,8), (3,7), (4,12), etc. in Fig. \ref{Fig:PWPconcept}.


{\it Matrix-vector multiplication} The measurement facilitated operation of matrix-vector multiplication follows from Fig. \ref{Fig:PWPconcept}. Suppose that $A_i$ is the desired product of the vector $J_i$ and the matrix $F_{ij}$. We rescale $J_j$ with a certain multiplier ($z_1$) to a convenient interval of electrode voltages ${\cal E}_j$. Secondly, using a proper multiplier ($z_2$), we rescale $F_{ij}$ so that all its elements fall in the interval of PWP system conductances,  $\delta G=\delta R/\langle R_{ij}^2\rangle$ with $\delta R$ from Eq. (\ref{eq:deltaR}).  The desired product becomes $A_i=z_1z_2I_i$ where $I_i=\sum _jG_{ij}{\cal E}_j$ is the current through the $i$th electrode in Fig. \ref{Fig:PWPconcept}. Because the conductance matrix $G_{ij}=R_{ij}^{-1}$ contains exponentially large number ($M\gg 1$) of elements closely covering interval $\delta G$, any desired value of $G_{ij}$ can be found among the measured conductances with good accuracy {\it without any additional actions}. After that, applying voltage ${\cal E}_j$ to the electrode $j$ produces a measurable current $G_{ij}{\cal E}_j$ through electrode $i$, which can be stored e. g. as a partial charge on a certain capacitor $C_i$. Measuring the total of all such partial contributions supplied by electrode $i$ in response to various ${\cal E}_j$ will give the component of sought vector $I_i$.

{\it Brain-like associative learning} commonly illustrated with Pavlov's dog salivation experiments (see e. g. Ref. \onlinecite{kuzum2013}) is readily implemented utilizing shared portions between bonds of a PWP cluster, such as bonds (1,5) and (1,7) in Fig. \ref{Fig:PWPconcept}. Identifying the `sight of food' and `sound' stimuli with signals on the electrodes 5 and 7, predicts that properly and simultaneously triggering both will switch their corresponding pathways (1,5) and (1,7) in a low resistive state making both salivation triggering (through the output on electrode 1).  In general, conductive pathways connecting various pairs of the electrodes and sharing the same portion of a PWP cluster will be mutually affected by a single bias-induced change demonstrating synapse functionality and a single-trial learning model for storage and retrieval of information resembling that of the cortex of the mammalian brain.

{\it Other functionalities} based on PCM and RRAM structures for neuromorphic computing are all attainable for PWP systems that retain their useful properties further amplifying them within a new architecture.

\section{PWP metrics}
We briefly mention several metrics \cite{kuzum2013} of the proposed PWP devices. (1) Dimensions and architecture: the above estimate of a superior information density may be reduced to account for larger physical dimension of a single microscopic resistor. However, even assuming $a$ in the range of tens of microns yields the density $\sim 10^9$ cm$^{-3}$, still above that of any existing memories; finite resolution of resistor reading can lower that figure. (2) Energy consumption. Assuming PWP structures with the same materials as PCM and RRAM, we expect its energy efficiency will be lower because of the lack of energy costly interconnects. (3) Operating speed/programming time. Generally, PWP devices RC times are greater than that of nano RRAM and PCM.  Like other brain-inspired systems, their computational efficiency will be achieved through parallelism. (4) Multi-level states: assuming $\sim$ 10 nm microscopic resistor and 1 cm device, each bond in a PWP cluster will contain hundreds of micro-resistors; hence, hundreds of multi-level states per typical bond, at the level allowing robust analog operations. (6) Retention and endurance: PWP systems will be superior to the existing PCM/RRAM based devices because of the lack of multiple interconnects triggering degradation.

\section{Conclusions}
We have introduced the percolation with plasticity systems that, while being essentially random, exhibit various neuromorphic functionalities and similarities to the cortex of mammalian brain. These systems demonstrate rich physics, understanding of which remains in its infancy. The outstanding challenges -in theory- go far beyond the standard percolation paradigm including statistical aspects, microscopic phase transformations, heat transfer and AC propagation in random systems with multiple co-existing interfaces. Both analytical research and numerical modeling will help to better understand PWP objects.  Multiple material bases can be used to fabricate PWP devices, ranging from the known PCM and RRAM materials to the nanocomposites insufficiently explored for memory and other neuromorphic applications.

\end{document}